\documentclass[10pt]{article}    
\usepackage{graphicx}

\usepackage{cite} 
\usepackage{url}  
\usepackage{ifthen}  
\usepackage{multicol}   
\usepackage[utf8]{inputenc} 

\newcommand{\doo}{\mathsf{do}}
\newcommand{\PP}{\mathbb{P}}
\newcommand{\EE}{\mathbb{E}}
\newcommand{\pa}{\mathsf{pa}}
\usepackage{bbm}
\usepackage{amsmath}
\usepackage{amsfonts}
\usepackage{amssymb}
\usepackage{lscape}

\usepackage{amsmath}
\usepackage{amsfonts}
\usepackage{amssymb}
\usepackage{multirow}

\begin{document}


\title{Joint estimation of causal effects from observational and intervention gene expression data}
 

\author{Andrea Rau$^{\star1,2}$, 
          Florence Jaffr\'ezic$^{1,2}$, 
         and 
          Gr\'egory Nuel$^3$%
     }


\date{}

\maketitle


\begin{scriptsize}
\indent    $^1$INRA, UMR1313 G{\'e}n{\'e}tique animale et biologie int{\'e}grative, 78352 Jouy-enJosas, France\\
\indent    $^2$AgroParisTech, UMR1313 G{\'e}n{\'e}tique animale et biologie int{\'e}grative, 75231 Paris 05, France\\
\indent    $^3$MAP5, UMR CNRS 8145, University Paris Descartes, 45 rue des Saints-P\`eres, F-75006 Paris - France\\
\\
\indent Email: Andrea Rau$^\star$ - andrea.rau@jouy.inra.fr; Florence Jaffr{\'e}zic - florence.jaffrezic@jouy.inra.fr; Gr{\'e}gory Nuel - \\
\indent gregory.nuel@parisdescartes.fr\\
\\
\indent $^\star$Corresponding author
\end{scriptsize}

\begin{abstract}
{\bf Background:}
Inference of gene regulatory networks from transcriptomic data has been a wide research area in recent years. Proposed methods are mainly based on the use of graphical Gaussian models for observational wild-type data and provide undirected graphs that are not able to accurately highlight the causal relationships among genes. 
In the present work, we seek to improve estimation of causal effects among genes by jointly modeling observational transcriptomic data with intervention data obtained by performing knock-outs or knock-downs on a subset of genes. By examining the impact of such expression perturbations on other genes, a more accurate reflection of regulatory relationships may be obtained than through the use of wild-type data alone.

{\bf Results:} 
Using the framework of Gaussian Bayesian networks, we propose a Markov chain Monte Carlo algorithm with a Mallows model and an analytical likelihood maximization to sample from the posterior distribution of causal node orderings, and in turn, to estimate causal effects.
The main advantage of the proposed algorithm over previously proposed methods is that it has the flexibility to accommodate any kind of intervention design, including partial or multiple knock-out experiments.
Methods were compared on simulated data as well as data from the Dialogue for Reverse Engineering Assessments and Methods (DREAM) 2007 challenge. 

{\bf Conclusions:} The simulation study confirmed the impossibility of estimating causal orderings of genes with observation data only. 
The proposed algorithm was found, in most cases, to perform better than the previously proposed methods in terms of accuracy for the estimation of causal effects.
In addition, multiple knock-outs proved to bring valuable additional information compared to single knock-outs. The choice of optimal intervention design therefore appears to be a crucial aspect for causal inference and an interesting challenge for future research.

\end{abstract}


\section*{Background}

The inference of gene regulatory networks from transcriptomic data has been a wide research area in recent years. Several approaches have been proposed for network inference from observational transcriptomic data (also referred to as wild-type or steady-state expression data), mainly based on the use of graphical Gaussian models \cite{glasso}. These methods, however, rely on the estimation of partial correlations and provide undirected graphs that cannot highlight the causal relationships among genes. Maathuis \textit{et al.} \cite{Buhlmann2010,Buhlmann2009} recently proposed a method called {\it Intervention-calculus when the DAG is Absent} (IDA) to predict bounds for causal effects from observational data alone in the context of Gaussian Bayesian networks (GBN). In the IDA, the PC-algorithm \cite{Kalisch2012,Kalisch2007} is first applied to find the associated completed partially directed acyclic graph (CPDAG), corresponding to the graphs belonging to the appropriate equivalence class. Following this step, bounds for total causal effects of each gene on the others are estimated using intervention calculus \cite{Pearl2000} for each directed acyclic graph (DAG) in the equivalence class.

However, if intervention experiments such as gene knock-outs or knock-downs are available, it is valuable to jointly perform causal network inference from a combination of wild-type and intervention data. One such approach has been proposed by Pinna \textit{et al.} \cite{Pinna2010}, based on the simple idea of calculating the deviation between observed gene expression values and the expression under each systematic intervention. 
In particular, Pinna \textit{et al.} propose the calculation of several matrices to evaluate the differences between observational and intervention expression values: a simple deviation matrix, a standardized deviation matrix, and a z-score deviation matrix.
In order to evaluate all possible causality links among genes, the method requires a single replicate of observational data as well as a single replicate of intervention data for each gene in the network. 


The method proposed in \cite{Pinna2010} has the advantage of being very fast to compute and is quite general, as it does not require any assumption of acyclicity of the graph. In addition, as this method provided the best network estimation in the Dialogue for Reverse Engineering Assessments and Methods (DREAM4) challenge \cite{stolovitzky2007}, it may be considered as a reference. However, we note that it requires an intervention experiment to be performed for each gene in the network, which is very constraining, and tends to provide very noisy estimations for the strength of causal effects.

To address these issues, the aim of this work is to propose a method using a  Markov chain Monte Carlo (MCMC) algorithm and Mallows model that is flexible enough to accurately infer causal gene networks from an arbitrary mixture of observational and intervention data, including partial and multiple gene knock-out experiments. As such, the proposed method is able to fully make use of all available information, does not require an intervention to be performed for each gene, and can deal with sophisticated multiple intervention designs. The proposed method was compared to those of \cite{Buhlmann2009} and \cite{Pinna2010} on simulated data as well as the data from the DREAM4 challenge \cite{stolovitzky2007}. 

\section*{Methods}

\subsection*{Gaussian Bayesian network framework}

Let $G=(V,E)$ be a graph defined by a set of vertices $V$ and edges $E \subset (V \times V)$. Let the vertices of a graph represent $p$ random variables $X_1,...,X_p$.
As in the approach of \cite{Buhlmann2009}, we consider here the framework of Gaussian Bayesian networks (GBN), which correspond to Bayesian networks where the nodes have a Gaussian residual distribution and edges represent linear dependencies. It also follows that in this case the joint distribution of the network is multivariate Gaussian.

In DAGs such as GBNs, we often encounter the presence of Markov equivalence classes, i.e. multiple network structures that yield the same joint distribution; in such cases, observational data alone generally cannot orient edges. For this reason, in many cases the use of intervention data can help overcome this issue, as presented below.

\subsubsection*{Calculation of total causal effects}
 
Following an intervention on a given node $X_i$, denoted as $\doo (X_i = x)$, we consider the expected value of each other gene in the network via do-calculus \cite{Pearl2000}:
\begin{align*}
\EE(X_j \vert \doo(X_i  = x)) =
	\begin{cases}
 	\EE(X_j) & \text{if } X_j \in \pa(X_i)\\
	\int \EE(X_j \vert x, \pa(X_i)) \PP(\pa(X_i)) d\pa(X_i) & \text{if} X_j  \notin \pa(X_i) \\
 	\end{cases}
\end{align*}
where $\pa(X_i)$ represents the parents of node $X_i$.
It is important to point out that $\PP(Y \vert \doo(X = x))$ is different from the conditional probability $\PP(Y \vert X = x)$. 
Using this framework, the total causal effects may be defined as follows:
$$ \beta_{ij} = \frac{\partial}{\partial x} \EE(X_j \vert \doo (X_i = x)) $$
and are equal to 0 if $X_i$ is not an ancestor of $X_j$. On the other hand, the direct causal effects (i.e. the edges in the graph) are defined as:
$$ \alpha_{ij} = \frac{\partial}{\partial x} \EE(X_j \vert \pa(X_j), \doo (X_i = x)). $$

%

\subsection*{Proposed causal inference method}

In the GBN framework, when observational data are jointly modeled with intervention data for an arbitrary subset of genes, for each sample the network follows a multivariate Gaussian distribution of dimension equal to the number of genes that had no intervention (as the expression value of the gene under intervention is fixed to a given value), and the log-likelihood value can subsequently be calculated for a proposed network. 

The calculations in the following section assume that the nodes in the graph have been sorted according to an appropriate causal ordering such that if $i < j$, then $X_j$ is not an ancestor of $X_i$; we note that such an ordering is possible under the assumption of acyclicity of the graph. In practice, of course, it is typically not possible to correctly order nodes in such a way without knowledge of the underlying DAG. For this reason, we aim to explore various network structures based on causal orderings, and to choose among those with the best likelihood value for an arbitrary set of observational and intervention data. The Metropolis-Hastings algorithm \cite{Metropolis53,Hastings70}, through the use of a proposal distribution for causal orderings, allows such an exploration to take place and to approach a local maximum of the likelihood.

\subsubsection*{Likelihood calculation}

Let $p$ be the number of nodes in the graph, $G$ the DAG structure and $\mathbf{W}$ the matrix containing the values for all direct causal effects. The nodes are assumed to have been sorted by parental order for $G$ in the matrix $\mathbf{W}$, i.e. if $i < j$, then $X_j$ is not an ancestor of $X_i$; under such an ordering, denoted $\mathcal{O}$, $\mathbf{W}$ is an upper triangular matrix. We note that an appropriate casual node ordering of a given graph is not necessarily unique. In the GBN framework, it is assumed that each node of $G$ has a residual Gaussian distribution, independently from the rest of the network. 
Let us consider 
a set of $p$ Gaussian random variables defined by:
$$X_j=m_j+\sum_{i \in \text{pa}(j)} w_{i,j} X_i + \varepsilon_j \quad \text{with} \quad \varepsilon_j \sim \mathcal{N}(0,\sigma_j^2)$$
where the $\varepsilon_j$ are assumed to be independent.
Given the causal ordering structure of the graph, the model parameters are $\theta=(m,\sigma,w)$ where $w_{i,j}$ is nonzero only on $(i,j)\in \mathcal{E}=\{i \in \text{pa}(j),j \in \mathcal{I} \}$, that is the edge set.

It is easy to see that this model is equivalent to $X_\mathcal{I} \sim \mathcal{N}( \boldsymbol{\mu} , \boldsymbol{\Sigma})$ with:
$$ \boldsymbol{\mu}=m  \mathbf{L} \quad\text{and}\quad \boldsymbol{\Sigma}= \mathbf{L}^T\text{diag}(\sigma^2)\mathbf{L}= \sum_{j \in \mathcal{I}} \sigma_j^2 \mathbf{L}^T e_j^T e_j \mathbf{L} $$
where $e_j$ is a null row-vector except for its $j^\text{th}$ term which is equal to $1$, and where  $\mathbf{L}=(\mathbf{I}-\mathbf{W})^{-1}=\mathbf{I}+\mathbf{W}+\ldots+\mathbf{W}^{p-1}$ with $\mathbf{W}=(w_{i,j})_{i,j \in \mathcal{I}}$. Note that the nilpotence of $\mathbf{W}$ is due to the fact that $w_{i,j}=0$ for all $i\geqslant j$.

The log-likelihood of the model given $N$ observations $x^k=(x^k_1,\ldots,x^k_p)$ ($1 \leqslant k \leqslant N$) is then:
\begin{equation}
\ell(m,\sigma,w)=-\frac{Np}{2} \log (2\pi)-N\sum_{j \in \mathcal{I}} \log (\sigma_j) 
-\frac{1}{2}\sum_{k=1}^N \sum_{j \in \mathcal{I}} \frac{1}{\sigma_j^2} (x^k_j - x^k\mathbf{W} e_j^T -m_j)^2.
\end{equation}
As shown in the appendix, analytical formulae can be obtained for the derivatives with respect to parameters $(m,\sigma,w)$.

The likelihood presented above only takes into account observational data. Let us now consider the case of an arbitrary mixture of observational and intervention data.
We assume that we perform an intervention on a subset $\mathcal{J} \subset \mathcal{I}=\{1,\ldots,p\}$ of variables by artificially fixing the level of the corresponding variables to a value (typically 0 in the case of knock-out experiments): $\text{do}(X_\mathcal{J}=x_\mathcal{J})$. The model is then obtained by assuming that all $w_{i,j}=0$ for $(i,j) \in \mathcal{E}$ and $j \in \mathcal{J}$; we denote the corresponding matrix $\mathbf{W}_\mathcal{J}$. We also assume that the variables $X_j$ for $j \in \mathcal{J}$ are fully deterministic. As before, the resulting model is hence Gaussian: $X_{\mathcal{I}} | \text{do}(X_\mathcal{J}=x_\mathcal{J}) \sim \mathcal{N}(\boldsymbol{\mu}_{\mathcal{J}}(x_\mathcal{J}),
\boldsymbol{\Sigma}_{\mathcal{J}})$ with
$$
\boldsymbol{\mu}_{\mathcal{J}}(x_\mathcal{J})=
\boldsymbol{\nu}_\mathcal{J}(x_\mathcal{J}) \mathbf{L}_{\mathcal{J}}
,\quad
\boldsymbol{\Sigma}_{\mathcal{J}}=  \sum_{j \notin \mathcal{J}} \sigma_j^2 \mathbf{L}_{\mathcal{J}}^T \mathbf{D}_j \mathbf{L}_{\mathcal{J}} 
,
$$
$$
\boldsymbol{\nu}_\mathcal{J}(x_\mathcal{J})e_j^T=\left\{
\begin{array}{ll}
x_j & \text{if $j \in \mathcal{J}$} \\
m_j & \text{otherwise} \\
\end{array}
\right.
\quad\text{and}\quad
\mathbf{L}_{\mathcal{J}}=(\mathbf{I}-\mathbf{W}_{\mathcal{J}})^{-1}=\mathbf{I}+\mathbf{W}_{\mathcal{J}}+\ldots+\mathbf{W}_{\mathcal{J}}^{p-1}.
$$

For the likelihood calculation, we consider $N$ data generated under $x^k=(x^k_1,\ldots,x^k_p)$ ($1 \leqslant k \leqslant N$) with intervention on $\mathcal{J}_k$ (where $\mathcal{J}_k=\emptyset$ means no intervention). We denote by $\mathcal{K}_j=\{k,j \notin \mathcal{J}_k\}$, and by $N_j=|\mathcal{K}_j|$ its cardinal. The log-likelihood of the model can then be written as:
\begin{equation}
\ell(m,\sigma,w)=-\frac{\log(2\pi)}{2}\sum_j N_j -\sum_{j} N_j \log (\sigma_j)
-\frac{1}{2} \sum_{k} \sum_{j \notin \mathcal{J}_k} \frac{1}{\sigma_j^2} (x^k_j-x^k\mathbf{W}e_j^T-m_j)^2 .\label{eqn:llmixture}
\end{equation}
As previously, it can be shown that analytical formulae can be obtained for maximum likelihood estimators of the parameters $(m,\sigma,w)$ (see the appendix).  

\subsubsection*{Proposed MCMC algorithm}

The Metropolis-Hastings algorithm \cite{Metropolis53,Hastings70} is a random walk over $\Omega$, the parameter space of the model.
It relies on an instrumental probability distribution $Q$ which defines the transition from position $X_t$ to a new position $X$. The probability of moving from state $X_t$ to the new state $X$ is defined by:
\begin{align}
P(X_{t+1}=X|X_t)=\min\left\{ \frac{\pi(X) Q(X_t,X)}{\pi(X_t)Q(X,X_t)},1  \right\} \label{eqn:acceptanceratio}
\end{align}
where $\pi(X)$ is the likelihood function. 

In order to propose a new causal node ordering $\mathcal{O}^\star$ from the previous ordering $\mathcal{O}$, we propose to make use of the Mallows model \cite{Mallows57}. Briefly, under this model, the density of a proposed causal ordering is defined as follows:
\begin{align}
P(\mathcal{O}^\star) &= P(\mathcal{O}^\star \vert \mathcal{O}, \phi) \nonumber \\
&= \frac{1}{Z}\phi^{d(\mathcal{O}^\star,\mathcal{O})} \nonumber
\end{align}
where $\phi \in (0, 1]$ is a fixed temperature parameter, $Z$ is a normalizing constant, and
$d(\cdot, \cdot)$ is a dissimilarity measure between $\mathcal{O}$ and $\mathcal{O}^\star$ based on the number of pairwise ranking disagreements. In addition, we remark that as the temperature parameter $\phi$ approaches zero, the Mallows model approaches a uniform distribution over all causal orderings, and if $\phi=1$, the model corresponds to a dirac distribution on the reference ordering $\mathcal{O}$. 
In the following, we will use a reparameterization of the temperature coefficient $\phi$ such that $\phi=\exp(-1/\eta)$, with $\eta > 0$.
Due to the symmetry of $d$, it is clear that $P(\mathcal{O}^\star \vert \mathcal{O}, \phi)=P(\mathcal{O} \vert \mathcal{O}^\star, \phi)$, which allows a simplification of the $Q$ terms in the acceptance ratio in Equation~(\ref{eqn:acceptanceratio}).

Proposals for causal node orderings using the aforementioned Mallows model may be obtained by sampling using a repeated insertion model as described in \cite{Doignon2004}. Based on this new proposal for the node ordering $\mathcal{O}^\star$, maximum likelihood estimators may be calculated for the model parameters $\theta = (m, \sigma, w)$ using the likelihood described in Equation~(\ref{eqn:llmixture}). Subsequently, the Metropolis-Hastings ratio may be calculated and used to determine whether the proposed causal node ordering is accepted or rejected. 


\begin{figure}[!t]
\centering
\includegraphics[width=.65\textwidth]{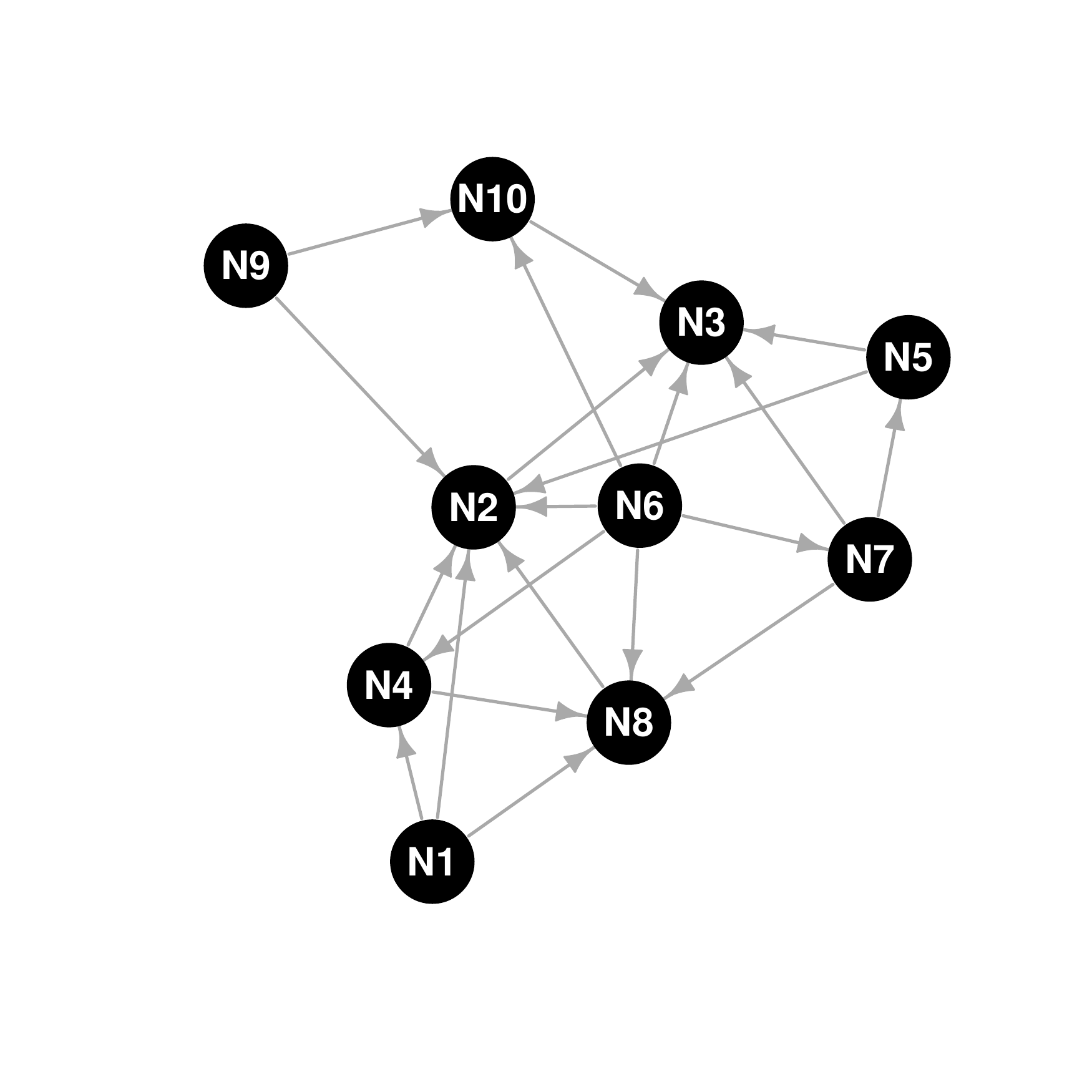}
\caption{{\bf Graph structure used in simulation study.} Graph structure taken from \cite{Kalisch2007} used for the simulation study for a graph with ten nodes and 21 edges.}
\end{figure}

\section*{Results and discussion}
\subsection*{Simulation study}

We considered simulated data from a network with 10 genes and 21 edges as described in \cite{Kalisch2007}; the underlying structure is given in Figure~1. 
For the residual distributions of each gene, we chose 0.5 for the means and three settings for the standard deviations ($\sigma=0.01$, 0.1 and 0.5), which correspond to little, moderate and large noise for the marginal distributions. Non-zero parameters $w_{ij}$ were simulated with values drawn uniformly from $(-1,-0.25) \cup (0.25,1)$, and 100 datasets were generated. The goal was to try to accurately infer the total causal effects present among genes. 

Several intervention designs were simulated: 1) 20 observational (wild-type) replicates with no interventions, 2) a mixed setting with 10 wild-types and one knock-out per gene, 3) a partial knock-out design with 15 wild-types and one knock-out for five genes \{1, 2, 3, 4, 5\}, and 4) a multiple knock-out design
with 10 wild types, one knock-out per gene and five double knock-outs: \{1,2\}, \{1,3\}, \{4,5\}, \{5,6\}, \{3,8\}.
Note that we have previously shown \cite{Nuel2013} that observational data alone (Setting 1 described above) are not informative for the causal node ordering as in such a case, the likelihood is invariant to permutations of the order. Consequently, in this setting node orderings were uniformly sampled rather than using the MCMC-Mallows algorithm; we refer to this strategy as MCMC-uniform.

An MCMC algorithm with Mallows proposal distribution was run to explore the posterior distribution of causal node orderings, as presented in the previous section, with full estimation of $\theta = (m, \sigma, w)$ using the maximum likelihood estimators. For the simulations, a small trial run of 1000 iterations was run over a range of possible temperature values $\eta$ (0.2 to 1.5 by 0.1) for the Mallows model, and the value yielding an acceptance rate closest to 30 to 40\% \cite{Roberts1997} was subsequently used for the full run of the MCMC algorithm. In all simulation settings tested here, this value was chosen to be $\eta=0.6$ (for $\sigma$ = 0.01 and 0.1) or $\eta=1$ (for $\sigma=0.5$). The MCMC-Mallows algorithm was subsequently run for 50,000 iterations, including a burn-in of 5000 iterations and thinning every 50 iterations. We note that due to the analytical maximization step of the likelihood, the method is quite fast and takes only a few minutes to run for each dataset.

The proposed algorithm was compared to two previously proposed methods: 1) Pinna \cite{Pinna2010}, which requires a very strict design with a knock-out for every gene, and 2) IDA \cite{Buhlmann2009} using the PC-algorithm, which can only deal with observational data.
As the PC-algorithm used by \cite{Buhlmann2009} provides bounds $(a,b)$ for the estimated causal effects, we considered two options to facilitate comparisons with the other methods: an ``optimistic" calculation, where we use the value $\max(\mathsf{abs}(a,b))$, and a more conservative ``pessimistic" strategy, using the value 
$\min(\mathsf{abs}(a,b))$ if $a$ and $b$ have the same sign, 0 otherwise.

Finally, several criteria were used to compare the different methods: area under the receiver operating characteristic (ROC) curve (AUROC), area under the precision-recall curve (AUPRC), Spearman correlation between true and estimated causal effects, and the mean squared error (MSE) of estimated causal effects.
Note that the results are calculated for the full $\mathbf{L}=(\mathbf{I}-\mathbf{W})^{-1}$ matrices (with the exception of the diagonal) and not just the upper triangular.

 \begin{table}[t!]
 \centering
 \caption{{\bf Comparison of methods for simulated data with moderate variability ($\sigma = 0.1$).} Several intervention designs were simulated: 1) 20 observational (wild-type) replicates with no interventions, 2) mixed setting with 10 wild-types and one knock-out per gene, 3) partial knock-out design with 15 wild-types and one knock-out for five genes \{1, 2, 3, 4, 5\}, and 4) multiple knock-out design with 10 wild types, one knock-out per gene and five double knock-outs: \{1,2\}, \{1,3\}, \{4,5\}, \{5,6\}, \{3,8\}.
Results were averaged over 100 simulations (standard deviations in parentheses): area under the ROC curve (AUROC), area under the precision-recall curve (AUPRC), Spearman correlation between true and estimated causal effects, and mean squared error (MSE) of estimated causal effects.}
     \begin{tabular}{cccccc}
Setting & Criterion       &  MCMC-Mallows & Pinna & IDA (opt) & IDA (pes) \\ \hline
\multirow{4}{*}{Observation only} &
        AUROC & 0.749 (0.043) & --- & 0.76 (0.062) & 0.643 (0.079) \\
        & AUPRC & 0.638 (0.053) & --- & 0.628 (0.078) & 0.527 (0.088) \\
        & Spearman & 0.48 (0.091) & --- & 0.491 (0.128) & 0.254 (0.177) \\
        & MSE & 0.056 (0.007) & --- & 0.182 (0.054) & 0.126 (0.034) \\
\hline
 \multirow{4}{*}{Mixed} &
        AUROC & 0.948 (0.03) & 0.825 (0.048) & 0.733 (0.068) & 0.67 (0.073) \\
        & AUPRC & 0.868 (0.042) & 0.737 (0.059) & 0.569 (0.087) & 0.53 (0.091) \\
        & Spearman & 0.696 (0.053) & 0.553 (0.097) & 0.42 (0.14) & 0.318 (0.186) \\
        & MSE & 0.026 (0.012) & 0.104 (0.011) & 0.334 (0.137) & 0.196 (0.067) \\
\hline
\multirow{4}{*}{Partial KO} &
        AUROC & 0.845 (0.059) & 0.795 (0.017) & 0.736 (0.056) & 0.646 (0.085) \\
        & AUPRC & 0.734 (0.078) & 0.725 (0.038) & 0.588 (0.075) & 0.514 (0.092) \\
        & Spearman & 0.587 (0.104) & 0.636 (0.034) & 0.449 (0.099) & 0.285 (0.187) \\
        & MSE & 0.035 (0.015) & 0.081 (0.008) & 0.215 (0.066) & 0.146 (0.049) \\
\hline
\multirow{4}{*}{Multiple KO} &
        AUROC & 0.959 (0.016) & 0.83 (0.035) & 0.733 (0.068) & 0.67 (0.073) \\
        & AUPRC & 0.886 (0.028) & 0.725 (0.039) & 0.569 (0.087) & 0.53 (0.091) \\
        & Spearman & 0.712 (0.028) & 0.625 (0.058) & 0.42 (0.14) & 0.318 (0.186) \\
        & MSE & 0.015 (0.006) & 0.107 (0.008) & 0.334 (0.137) & 0.196 (0.067) \\
\hline
    \end{tabular}
 \end{table}

Results are presented in Table~1 for $\sigma=0.1$, and in Supplementary Tables~1 and 2 in the appendix for $\sigma=0.01$ and $0.5$.
It can first be noted that results for the IDA method are identical for different levels of variation $\sigma$; this is due to the fact that it operates on sufficient statistics (correlation matrices) rather than on the data themselves. Similarly, results are identical for the MCMC-uniform method at different levels of $\sigma$ when only observational data are present.
Based on observational data only, we note that the proposed algorithm performs as well as the IDA approach; this is unsurprising as both methods are based on GBNs. 

When single knock-outs were simulated (one for each gene), for a large variability ($\sigma=0.5$), the performance of the IDA \cite{Buhlmann2009} approach is slightly better than Pinna \cite{Pinna2010} for accurate estimation of causal effects, although we recall that the former method solely makes use of the observational data.
On the other hand, when the amount of variability decreases ($\sigma=0.1$ and 0.01), the Pinna approach outperforms IDA, even for the optimistic version. In all three settings ($\sigma$ = 0.5, 0.1, 0.01), the proposed MCMC-Mallows algorithm was better able to estimate the causal effects than either Pinna or IDA, as shown by the different criteria presented here. A similar conclusion is obtained in the context of partial intervention design. The MCMC-Mallows approach was found to outperform the IDA approach, especially for moderate and low variability. The Pinna approach can unfortunately not be applied in this case as it requires systematic knock-outs to be performed. Finally, it was found that the addition of multiple knock-outs allowed an improvement of the estimation of the causal effects over single knock-outs alone. We note that this complex intervention design can only be accommodated by the proposed MCMC-Mallows method. In this setting, the Pinna method uses only information on the 10 single knock-outs and  the IDA approach only the observational data.

Figure~2 presents the posterior distribution of causal node ordering from the MCMC-Mallows method averaged over 100 simulations for the observation data only (left), the mixed setting with 10 wild types and one knock-out for each gene (middle), and the partial knock-out setting (right) for moderately noisy data ($\sigma = 0.1$). The node order distribution for the multiple knock-out design (see Supplementary Figure~4 in the appendix) was found to be very similar to the mixed setting. We may remark on several points. First, as shown in the Methods section, it is not possible to estimate the node orders from observation data only. 
As expected, the node orders were most accurately estimated when a complete knock-out design was considered, with one knock-out for each gene, than for a partial knock-out design. For low to medium variability ($\sigma=0.01$ and 0.1) the proposed algorithm was able to very accurately estimate the potential node orders for the complete and multiple knock-out designs (see Supplementary Figures 1-4 in the appendix). Finally, we note that the node ordering is not unique for the DAG considered here, as illustrated by the black squares in Figure~2.

\subsection*{DREAM data analysis}

The proposed MCMC-Mallows algorithm as well as the two previously presented methods \cite{Pinna2010,Buhlmann2009} were applied to data from the DREAM4 challenge, an international competition held yearly to contribute to the development of powerful inference methods \cite{stolovitzky2007}.
In the DREAM4 {\it in silico} network challenge, network topologies (with feedback loops) were extracted from transcriptional regulatory networks of {\it E. coli} and {\it S. cerevisiae}, and data were subsequently simulated and distributed to the participants. The goal was to infer directed regulatory networks from simulated data with either 10 or 100 genes. Based on the considered evaluation criteria (AUROC and AUPRC), the Pinna method \cite{Pinna2010} was found to be the best performer for the 100-gene network challenge.
In this paper we will focus on the five simulated 10-gene networks and perform inference based on wild type, knock-out and multifactorial perturbation data.

Figure~3 presents the ROC curves as well as the precision-recall curves for the different methods in each of the five DREAM4 datasets. It can first be observed that the IDA\cite{Buhlmann2009}, whether optimistic or pessimistic versions of the causal effects estimations are used, performs the worst; this is unsurprising, as it only makes use of the observational data. On the other hand, the proposed MCMC-Mallows method compares quite well to the Pinna approach, except for the first data set where Pinna clearly outperforms the others. On the other hand, the MCMC-Mallows algorithm performs better for the second and fifth data sets. 

We note that the simulated intervention setting was well adapted to the Pinna method, as one knock-out was available for each gene and feedbac loops were included in the graph. This method would, however, not be able to deal with a partial or more complex multiple knock-out design, as shown in the Simulation section above. Its practical application is therefore quite limited.

\begin{landscape}
\begin{figure}[p]
\centering
\includegraphics[height=.375\textwidth]{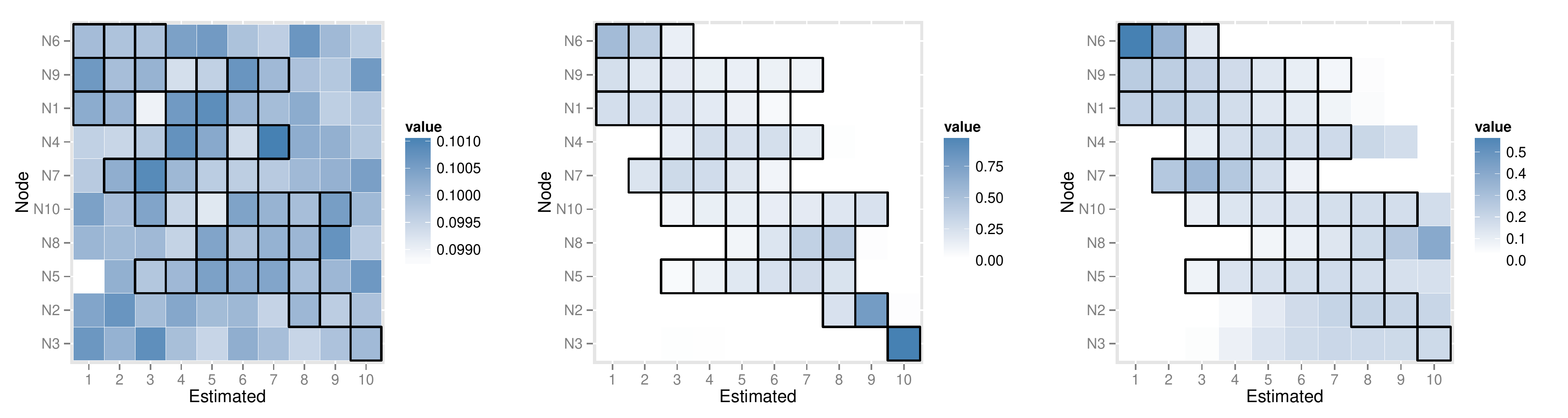}
\caption{{\bf Posterior distribution of node orders from the MCMC-Mallows approach, averaged over 100 simulations.} Results from simulation setting with $\sigma = 0.1$: Observations only (left), complete single knock-outs (middle), partial single knock-outs (right). The true node order (1 to 10) is represented in the rows, the estimated node orders in the columns, and the intensity of color of each square corresponds to the average proportion of iterations in which a given node was placed in a given position. As the causal node ordering is not unique for this DAG, true potential positions for each node are outlined in black.}
\end{figure}
\end{landscape}

\begin{figure}[!t]
\centering
\includegraphics[height=8cm]{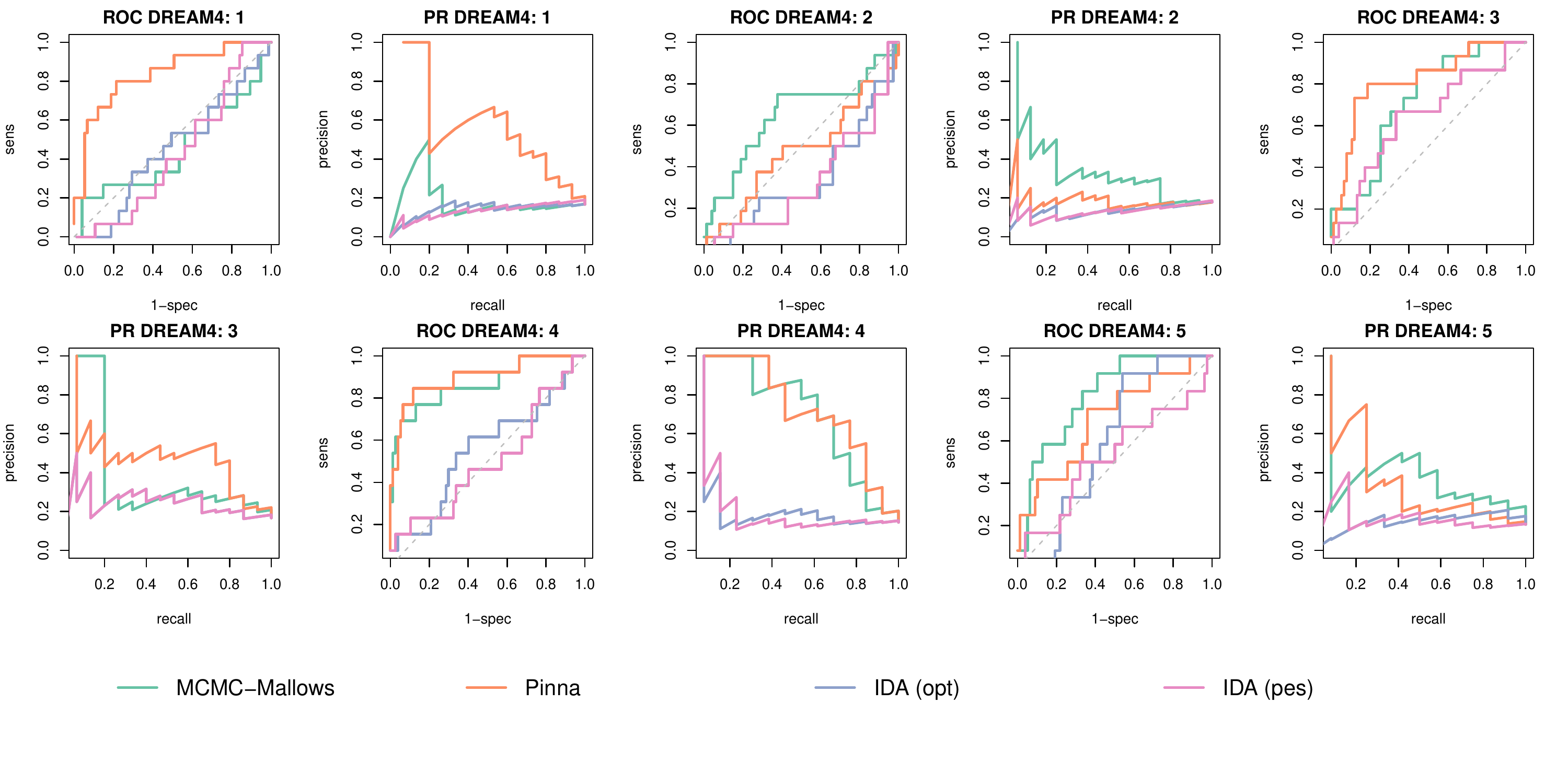}
\caption{{\bf Comparison of methods on data from the DREAM4 challenge.} ROC curves and precision-recall curves for the five simulated 10-gene networks of the DREAM4 challenge \cite{stolovitzky2007} for the MCMC-Mallows, Pinna, and IDA (optimistic and pessimistic) methods.}
\end{figure}

\section*{Conclusions}

In this paper we proposed a flexible and powerful approach for joint causal network inference from both observational and intervention data, using an MCMC algorithm and Mallows model. The computational efficiency of the method is very much improved by the analytical maximization step of the likelihood. This method has several advantages compared to the previously proposed methods. In particular, on the contrary to the Pinna approach \cite{Pinna2010}, which requires a very stringent design with one knock-out for each gene, it can deal with any knock-out designs and even with multiple knock-outs.

In the simulation study presented above, the proposed MCMC-Mallows algorithm was found to perform better than Pinna \cite{Pinna2010} and IDA \cite{Buhlmann2009} in terms of accuracy of estimation of the causal effects, as evidenced by the tendancy to have larger AUROC, larger Spearman correlation coefficients and smaller MSE than the other approaches.
Multiple knock-out designs were also found to bring more information for causal network inference than single knock-outs.

These results suggest that the choice of optimal experimental knock-out designs is a critical aspect for causal inference and merits further attention. Hauser and B\"uhlmann \cite{Hauser2012} recently proposed two strategies for the choice of optimal interventions. The first is a greedy approach using single-vertex interventions that maximizes the number of edges that can be oriented after each intervention; the second yields a minimum set of targets of arbitrary size that guarantee full identifiability. However, alternative approaches could be envisaged in future research. In particular, recall that in the GBN framework, the likelihood associated to the multivariate Gaussian distribution of the network can be explicitly written as presented in this work. The choice of optimal knock-outs to be performed to improve and validate the causal inference can then rely on the evaluation of the amount of information contributed by each possible intervention, which can for example be evaluated by the Fisher information. Its calculation requires the derivation of the likelihood function, which is not trivial but has already been derived in \cite{Nuel2013}. We anticipate that this issue will remain an interesting challenge for future research.

\bigskip

\section*{Authors' contributions}
AR participated in the design of the study, performed simulations and data analyses, and helped draft the manuscript. FJ participated in the design of the study and drafted the manuscript. GN designed the study, performed the analytical likelihood calculations and helped draft the manuscript.
All authors read and approved the final manuscript.   

\section*{Acknowledgements}
We thank R\'emi Bancal for his work during his master internship.

 \bibliographystyle{plain}  
 \bibliography{Intervention}

\begin{thebibliography}{10}

\bibitem{Doignon2004}
JP~Doignon, A~Pekec, and M~Regenwetter.
\newblock The repeated insertion model for rankings: Missing link between two
  subset choice models.
\newblock {\em Psychometrika}, 69(1):33--54, 2004.

\bibitem{glasso}
J~Friedman, T~Hastie, and R~Tibshirani.
\newblock {Sparse inverse covariance estimation with the graphical lasso}.
\newblock {\em Biostatistics}, 9(3):432--441, July 2008.

\bibitem{Hastings70}
WK~Hastings.
\newblock {Monte Carlo sampling methods using Markov chains and their
  applications}.
\newblock {\em Biometrika}, 57(1):97--109, April 1970.

\bibitem{Hauser2012}
A~Hauser and P~B\"uhlmann.
\newblock Two optimal strategies for active learning of causal models from
  interventions.
\newblock {\em Proc. of the 6th European Workshop on Probabilistic Graphical
  Models}, pages 123--130, 2012.

\bibitem{Kalisch2007}
M~Kalisch and P~B\"{u}hlmann.
\newblock Estimating high-dimensional directed acyclic graphs with the
  pc-algorithm.
\newblock {\em J. Mach. Learn. Res.}, 8:613--636, May 2007.

\bibitem{Kalisch2012}
M~Kalisch, M~M\"achler, D~Colombo, MH~Maathuis, and P~B\"uhlmann.
\newblock Causal inference using graphical models with the r package pcalg.
\newblock {\em Journal of Statistical Software}, 47(11):1--26, 5 2012.

\bibitem{Buhlmann2010}
MH~Maathuis, D~Colombo, M~Kalisch, and P~B\"uhlmann.
\newblock Predicting causal effects in large-scale systems from observational
  data.
\newblock {\em Nature Methods}, 7:247--248, 2010.

\bibitem{Buhlmann2009}
MH~Maathuis, M~Kalisch, and P~B\"{u}hlmann.
\newblock {Estimating high-dimensional intervention effects from observational
  data}.
\newblock {\em Annals of Statistics}, 37:3133--3164, 2009.

\bibitem{Mallows57}
CL~Mallows.
\newblock Non-null ranking models.
\newblock {\em Biometrika}, 44:114--130, 1957.

\bibitem{Metropolis53}
N~Metropolis, AW~Rosenbluth, MN~Rosenbluth, AH~Teller, and E~Teller.
\newblock {Equations of state calculations by fast computing machines}.
\newblock {\em Journal of Chemical Physics}, 21(6):1087--1092, 1953.

\bibitem{Nuel2013}
G~Nuel, A~Rau, and F~Jaffr{\'e}zic.
\newblock Joint likelihood calculation for intervention and observational data
  from a gaussian bayesian network.
\newblock {\em arXiv:1305.0709v4}.

\bibitem{Pearl2000}
J~Pearl.
\newblock {\em {Causality: Models, Reasoning and Inference}}.
\newblock Cambridge University Press, New York, NY, USA, 2000.

\bibitem{Pinna2010}
A~Pinna, N~Soranzo, and A~de~la Fuente.
\newblock From knockouts to networks: establishing direct cause-effect
  relationships through graph analysis.
\newblock {\em PLoS ONE}, 10(5):e12912, 2010.

\bibitem{Roberts1997}
GO~Roberts, A~Gelman, and WR~Gilks.
\newblock {Weak convergence and optimal scaling of random walk Metropolis
  algorithms}.
\newblock {\em Annals of Applied Probability}, 7:110--120, 1997.

\bibitem{stolovitzky2007}
G~Stolovitzky, D~Monroe, and A~Califano.
\newblock {Dialogue on Reverse-Engineering Assessment and Methods : The DREAM
  of high-throughput pathway inference}.
\newblock {\em Ann N Y Acad Sci}, 1115:1--22, 2007.

\end{thebibliography}

\appendix
\setcounter{table}{0}
\setcounter{figure}{0}
\renewcommand{\figurename}{Supplementary Figure}
\renewcommand{\tablename}{Supplementary Table}

\section{Additional details of analytical formulae for maximum likelihood parameter estimation }

\subsection{Observational data}

\subsubsection*{Likelihood calculation}

The log-likelihood of the model given $N$ observations $x^k=(x^k_1,\ldots,x^k_p)$ ($1 \leqslant k \leqslant N$) is:
\begin{equation}
\ell(m,\sigma,w)=-\frac{Np}{2} \log (2\pi)-N\sum_{j \in \mathcal{I}} \log (\sigma_j) 
-\frac{1}{2}\sum_{k=1}^N \sum_{j \in \mathcal{I}} \frac{1}{\sigma_j^2} (x^k_j - x^k\mathbf{W} e_j^T -m_j)^2
\label{eq:likelihood_obs}
\end{equation}
Proof: for all $k$, let us define $A_k= (x^k-m\mathbf{L}) \boldsymbol{\Sigma}^{-1} (x^k-m\mathbf{L})^T$. Since
$\boldsymbol{\Sigma}^{-1}=(\mathbf{I}-\mathbf{W})\text{diag}(1/\sigma^2)(\mathbf{I}-\mathbf{W})^T$ we get:
\begin{eqnarray*}
A_k &=& \sum_{j \in \mathcal{I}} \frac{1}{\sigma_j^2} (x^k(\mathbf{I}-\mathbf{W})-m)e_j^T e_j (x^k(\mathbf{I}-\mathbf{W})-m)^T\\
&=&\sum_{j \in \mathcal{I}} \frac{1}{\sigma_j^2} (x^k_j - x^k\mathbf{W} e_j^T -m_j)^2.
\end{eqnarray*}

\subsubsection*{Derivatives with respect to $m$}

$$
\frac{\partial \ell}{\partial m_j}(\theta)
= \frac{1}{\sigma_j^2} \sum_{k=1}^N  (x^k_j -x^k\mathbf{W}e_j^T -m_j)
$$

The maximization of $\ell(\theta)$ in $m$ for a fixed $w$ does not depend on $\sigma$ and is given by:
\begin{equation}
m_j=\frac{1}{N} \sum_{k=1}^N (x^k_j -x^k\mathbf{W}e_j^T)
\end{equation}
by replacing $m_j$ by this formula in Eq.~\ref{eq:likelihood_obs} we get an expression of the likelihood free of the parameter $m$:
\begin{equation}
\tilde{\ell}(\sigma,w)=-\frac{Np}{2} \log (2\pi)-N\sum_{j \in \mathcal{I}} \log (\sigma_j) 
-\frac{1}{2}\sum_{k=1}^N \sum_{j \in \mathcal{I}} \frac{1}{\sigma_j^2} (y^k_j - y^k\mathbf{W} e_j^T)^2
\end{equation}
with $y^k_j=x^k_j-1/N\sum_{k'} x^{k'}_j$ for all $k,j$.

\subsubsection*{Derivatives with respect to $w$}

$$
\frac{\partial \ell}{\partial w_{i,j}}(\theta)=
\frac{1}{\sigma_{j}^2}\sum_{k=1}^N y^k_i (y^k_j- y^k\mathbf{W}e_j^T).
$$
Proof:
$$
\sum_{j' \in \mathcal{I}} \frac{1}{\sigma_{j'}^2} (y^k_{j'} - y^k\mathbf{W} e_{j'}^T)(\underbrace{y^ke_i^T}_{y^k_i}\underbrace{e_j e_{j'}^T}_{\mathbbm{1}_{j'=j}})
=\frac{y_j^k}{\sigma_{j}^2} (y^k_{j} - y^k\mathbf{W} e_{j}^T)
$$

The maximization of $\tilde{\ell}(\sigma,w)$ in $w$ can then be done independently from $\sigma$  by solving for all $(i,j)\in \mathcal{E}$:
$$
\sum_{k=1}^N y^k_i y^k\mathbf{W}e_j^T= \sum_{k=1}^N y^k_i y^k_i
$$
hence using
$$
\mathbf{W}=\sum_{(i',j')\in \mathcal{E}} w_{i',j'} e_{i'}^T e_{j'}
\Rightarrow
y^k\mathbf{W}e_j^T = \sum_{i', (i',j)\in \mathcal{E}} w_{i',j} y^k_{i'} 
$$
so that we get:
\begin{equation}
\sum_{i', (i',j)\in \mathcal{E}} w_{i',j}\sum_{k=1}^N y_i^ky_{i'}^k = \sum_{k=1}^N y_i^k y_j^k
\quad\text{for all $(i,j)\in \mathcal{E}$}
\end{equation}

\subsubsection*{Derivatives with respect to $\sigma$}

$$
\frac{\partial \ell}{\partial \sigma_j}(\theta)=-\frac{N}{\sigma_j}
+\frac{1}{\sigma_j^3}\sum_{k=1}^N (y^k_j - y^k\mathbf{W}e_j^T)^2
$$

The maximization of $\tilde{\ell}(\sigma,w)$ in $\sigma$ when $m$ is fixed is then given by:
\begin{equation}
\sigma_j^2 
= \frac{1}{N} \sum_{k=1}^N (y^k_j - y^k\mathbf{W}e_j^T)^2
\end{equation}

\subsection{Mixture of observational and intervention data}

We consider $N$ data generated under $x^k=(x^k_1,\ldots,x^k_p)$ ($1 \leqslant k \leqslant N$) with intervention on $\mathcal{J}_k$ ($\mathcal{J}_k=\emptyset$ means no intervention). We denote by $\mathcal{K}_j=\{k,j \notin \mathcal{J}_k\}$, and by $N_j=|\mathcal{K}_j|$ its cardinal. The log-likelihood of the model can then be written as:
\begin{equation}
\ell(m,\sigma,w)=-\frac{\log(2\pi)}{2}\sum_j N_j -\sum_{j} N_j \log (\sigma_j)
-\frac{1}{2} \sum_{k} \sum_{j \notin \mathcal{J}_k} \frac{1}{\sigma_j^2} (x^k_j-x^k\mathbf{W}e_j^T-m_j)^2
\end{equation}
Proof: This is mainly due to the fact that for any intervention set $\mathcal{J}$ we have $\mathbf{W}_{\mathcal{J}}e_j^T=\mathbf{W}e_j^T$ for all $j \notin \mathcal{J}$.

\vspace{1em}

Considering the derivative with respect to $m_j$ we get for all $j$ such as $N_j>0$:
\begin{equation}
m_j=\frac{1}{N_j} \sum_{k \in \mathcal{K}_j} (x^k_j-x^k\mathbf{W}e_j^T)
\end{equation}
which can be plugged into the likelihood expression to get:
\begin{equation}
\tilde{\ell}(\sigma,w)=-\frac{\log(2\pi)}{2}\sum_j N_j-\sum_{j} N_j \log (\sigma_j)
-\frac{1}{2} \sum_{k} \sum_{j \notin \mathcal{J}_k} \frac{1}{\sigma_j^2} (y^{k,j}_j-y^{k,j}\mathbf{W}e_j^T)^2
\end{equation}
where for $(k,j)$ such as $j \notin \mathcal{J}_k$ we have:
$$
y^{k,j}=x^k-\frac{1}{N_j} \sum_{k'\in \mathcal{K}_j} x^{k'}
$$
and $w$ can be estimated by solving the following linear system:
\begin{equation}
\sum_{i', (i',j)\in \mathcal{E}} w_{i',j}\sum_{k \in \mathcal{K}_j} y_i^{k,j}y_{i'}^{k,j} = \sum_{k \in \mathcal{K}_j} y_i^{k,j} y_j^{k,j}
\quad\text{for all $(i,j)\in \mathcal{E}$}
\end{equation}
Note that the system might be degenerated if the intervention design gives no insight on some parameters.

It is hence finally possible to obtain $\sigma$ through:
\begin{equation}
\sigma_j^2 
= \frac{1}{N_j} \sum_{k \in \mathcal{K}_j} (y^{k,j}_j - y^{k,j}\mathbf{W}e_j^T)^2
\end{equation}

\clearpage
\section{Comparison of methods for simulated data with large variability ($\sigma = 0.5$) and small variability ($\sigma = 0.01$)}

\begin{table}[htbp]
    \centering
    \begin{tabular}{cccccc}
Setting      &  Criterion  & MCMC-Mallows & Pinna & IDA (opt) & IDA (pes) \\ \hline
\multirow{4}{*}{Observation only} &
        AUROC & 0.749 (0.043) & --- & 0.76 (0.062) & 0.643 (0.079) \\
        & AUPRC & 0.638 (0.053) & --- & 0.628 (0.078) & 0.527 (0.088) \\
        & Spearman & 0.48 (0.091) & --- & 0.491 (0.128) & 0.254 (0.177) \\
        & MSE & 0.056 (0.007) & --- & 0.182 (0.054) & 0.126 (0.034) \\ 
        \hline
 \multirow{4}{*}{Mixed} &
        AUROC & 0.791 (0.09) & 0.625 (0.064) & 0.733 (0.068) & 0.67 (0.073) \\
        & AUPRC & 0.654 (0.116) & 0.43 (0.069) & 0.569 (0.087) & 0.53 (0.091) \\
        & Spearman & 0.505 (0.145) & 0.181 (0.159) & 0.42 (0.14) & 0.318 (0.186) \\
        & MSE & 0.069 (0.027) & 0.553 (0.13) & 0.334 (0.137) & 0.196 (0.067) \\ 
        \hline
\multirow{4}{*}{Partial KO} &
        AUROC & 0.732 (0.072) & 0.731 (0.027) & 0.736 (0.056) & 0.646 (0.085) \\
        & AUPRC & 0.595 (0.097) & 0.545 (0.066) & 0.588 (0.075) & 0.514 (0.092) \\
        & Spearman & 0.431 (0.149) & 0.485 (0.071) & 0.449 (0.099) & 0.285 (0.187) \\
        & MSE & 0.063 (0.02) & 0.272 (0.073) & 0.215 (0.066) & 0.146 (0.049) \\ 
\hline
\multirow{4}{*}{Multiple KO} &
        AUROC & 0.819 (0.09) & 0.613 (0.07) & 0.733 (0.068) & 0.67 (0.073) \\
        & AUPRC & 0.7 (0.109) & 0.407 (0.077) & 0.569 (0.087) & 0.53 (0.091) \\
        & Spearman & 0.542 (0.128) & 0.162 (0.169) & 0.42 (0.14) & 0.318 (0.186) \\
        & MSE & 0.054 (0.022) & 0.399 (0.1) & 0.334 (0.137) & 0.196 (0.067) \\ 
        \hline
    \end{tabular}
\caption{{\bf $\sigma = 0.5$}. Several intervention designs were simulated: 1) 20 observational (wild-type) replicates with no interventions, 2) mixed setting with 10 wild-types and one knock-out per gene, 3) partial knock-out design with 15 wild-types and one knock-out for five genes \{1, 2, 3, 4, 5\}, and 4) multiple knock-out design with 10 wild types, one knock-out per gene and five double knock-outs: \{1,2\}, \{1,3\}, \{4,5\}, \{5,6\}, \{3,8\}. Results were averaged over 100 simulations (standard deviations in parentheses): area under the ROC curve (AUROC), area under the precision-recall curve (AUPRC), Spearman correlation between true and estimated causal effects, and mean squared error (MSE) of estimated causal effects.}
\end{table}

\begin{table}[tbp]
    \centering
    \begin{tabular}{cccccc}
Setting & Criterion         & MCMC-Mallows & Pinna & IDA (opt) & IDA (pes)  \\ \hline
\multirow{4}{*}{Observation only} &
        AUROC & 0.749 (0.043) & --- & 0.76 (0.062) & 0.643 (0.079) \\
        & AUPRC & 0.638 (0.053) & --- & 0.628 (0.078) & 0.527 (0.088) \\
        & Spearman & 0.48 (0.091) & --- & 0.491 (0.128) & 0.254 (0.177) \\
        & MSE & 0.056 (0.007) & --- & 0.182 (0.054) & 0.126 (0.034) \\ 
\hline
 \multirow{4}{*}{Mixed} &
        AUROC & 0.983 (0.015) & 0.945 (0.026) & 0.733 (0.068) & 0.67 (0.073) \\
        & AUPRC & 0.933 (0.017) & 0.902 (0.023) & 0.569 (0.087) & 0.53 (0.091) \\
        & Spearman & 0.744 (0.027) & 0.693 (0.046) & 0.42 (0.14) & 0.318 (0.186) \\
        & MSE & 0.001 (0.001) & 0.088 (0.001) & 0.334 (0.137) & 0.196 (0.067) \\ 
\hline
\multirow{4}{*}{Partial KO} &
        AUROC & 0.904 (0.032) & 0.829 (0.008) & 0.736 (0.056) & 0.646 (0.085) \\
        & AUPRC & 0.798 (0.058) & 0.803 (0.013) & 0.588 (0.075) & 0.514 (0.092) \\
        & Spearman & 0.645 (0.035) & 0.689 (0.017) & 0.449 (0.099) & 0.285 (0.187) \\
        & MSE & 0.016 (0.009) & 0.073 (0.001) & 0.215 (0.066) & 0.146 (0.049) \\ 
\hline
\multirow{4}{*}{Multiple KO} &
        AUROC & 0.986 (0.007) & 0.896 (0.009) & 0.733 (0.068) & 0.67 (0.073) \\
        & AUPRC & 0.937 (0.012) & 0.792 (0.006) & 0.569 (0.087) & 0.53 (0.091) \\
        & Spearman & 0.751 (0.015) & 0.691 (0.004) & 0.42 (0.14) & 0.318 (0.186) \\
        & MSE & 0.001 (0.001) & 0.097 (0.001) & 0.334 (0.137) & 0.196 (0.067) \\ 
\hline
 \end{tabular}
\caption{{\bf $\sigma = 0.01$}. Several intervention designs were simulated: 1) 20 observational (wild-type) replicates with no interventions, 2) mixed setting with 10 wild-types and one knock-out per gene, 3) partial knock-out design with 15 wild-types and one knock-out for five genes \{1, 2, 3, 4, 5\}, and 4) multiple knock-out design with 10 wild types, one knock-out per gene and five double knock-outs: \{1,2\}, \{1,3\}, \{4,5\}, \{5,6\}, \{3,8\}. Results were averaged over 100 simulations (standard deviations in parentheses): area under the ROC curve (AUROC), area under the precision-recall curve (AUPRC), Spearman correlation between true and estimated causal effects, and mean squared error (MSE) of estimated causal effects.}
\end{table}


\begin{landscape}
\begin{figure}[p]
\centering
\includegraphics[width=.40\textheight]{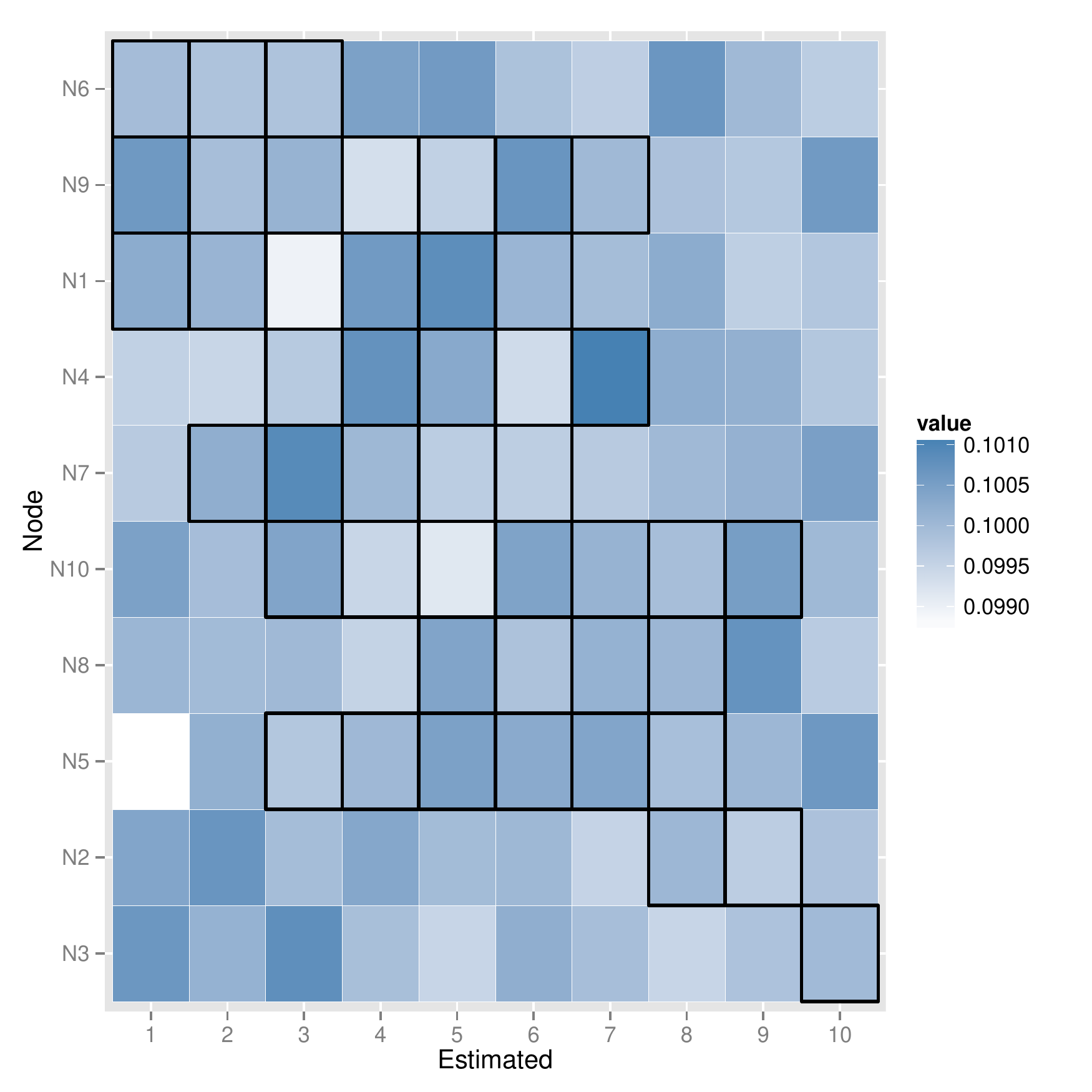}
\includegraphics[width=.40\textheight]{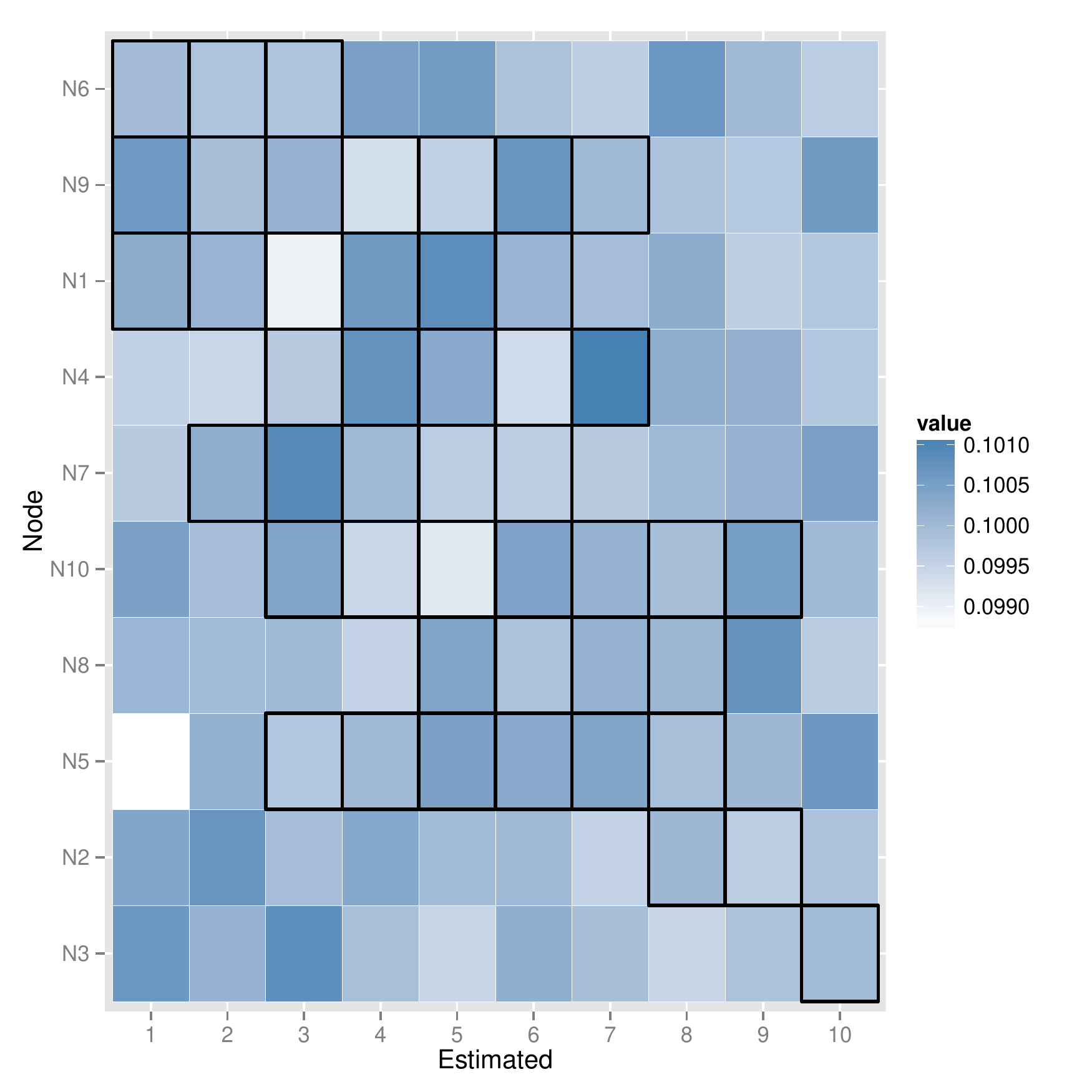}
\includegraphics[width=.40\textheight]{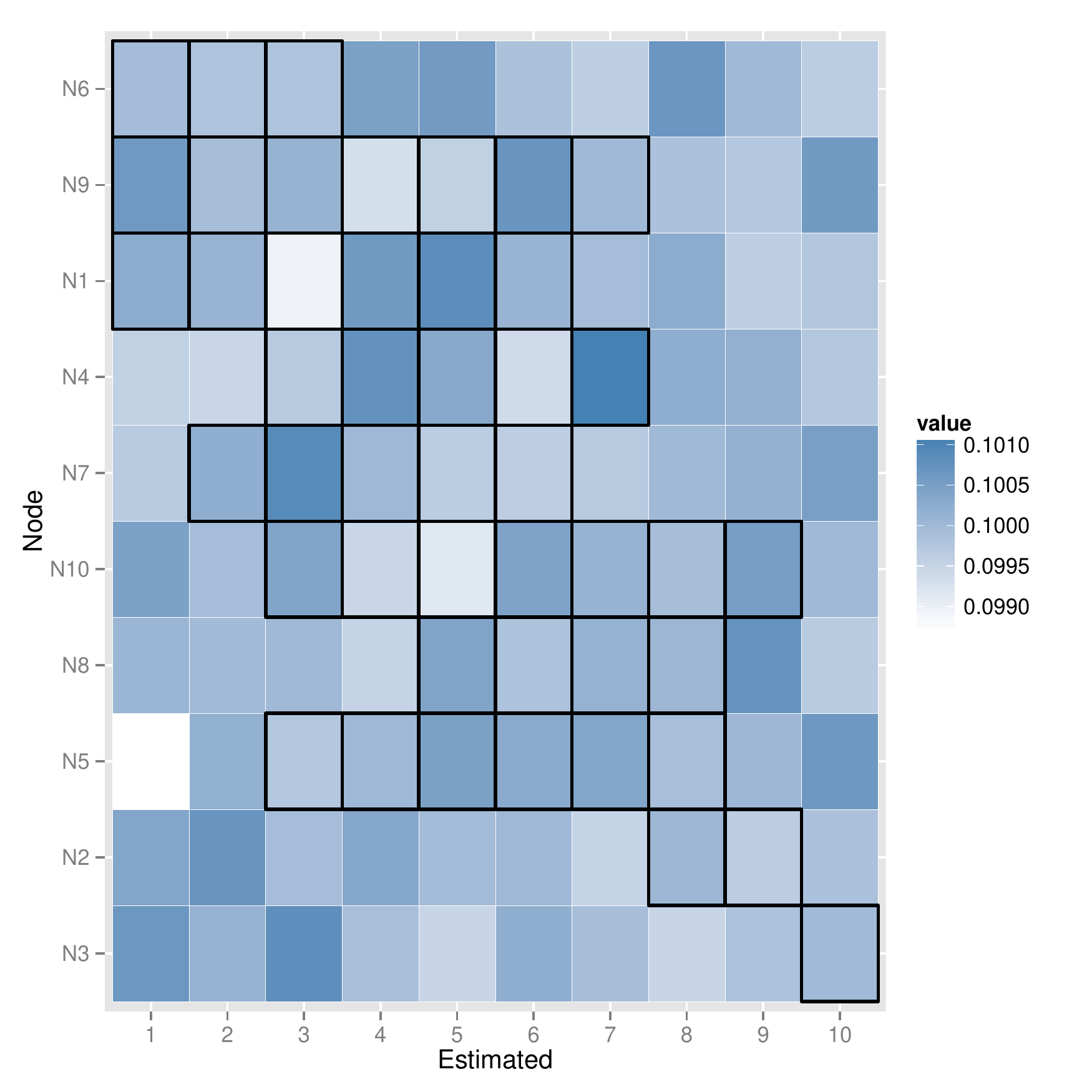}
\caption{{\bf $\sigma = 0.5$ (left), $\sigma = 0.1$ (middle), $\sigma = 0.01$ (right), observation only setting}. Posterior distribution of node orders, averaged over 100 simulations. The true node order (1 to 10) is represented in the rows, the estimated node orders in the columns, and the color of each square corresponds to the average proportion of iterations in which a given node was placed in a given position. As the node ordering is not unique for this DAG, true potential positions for each node are outlined in black.}
\end{figure}
\end{landscape}

\begin{landscape}
\begin{figure}[p]
\centering
\includegraphics[width=.40\textheight]{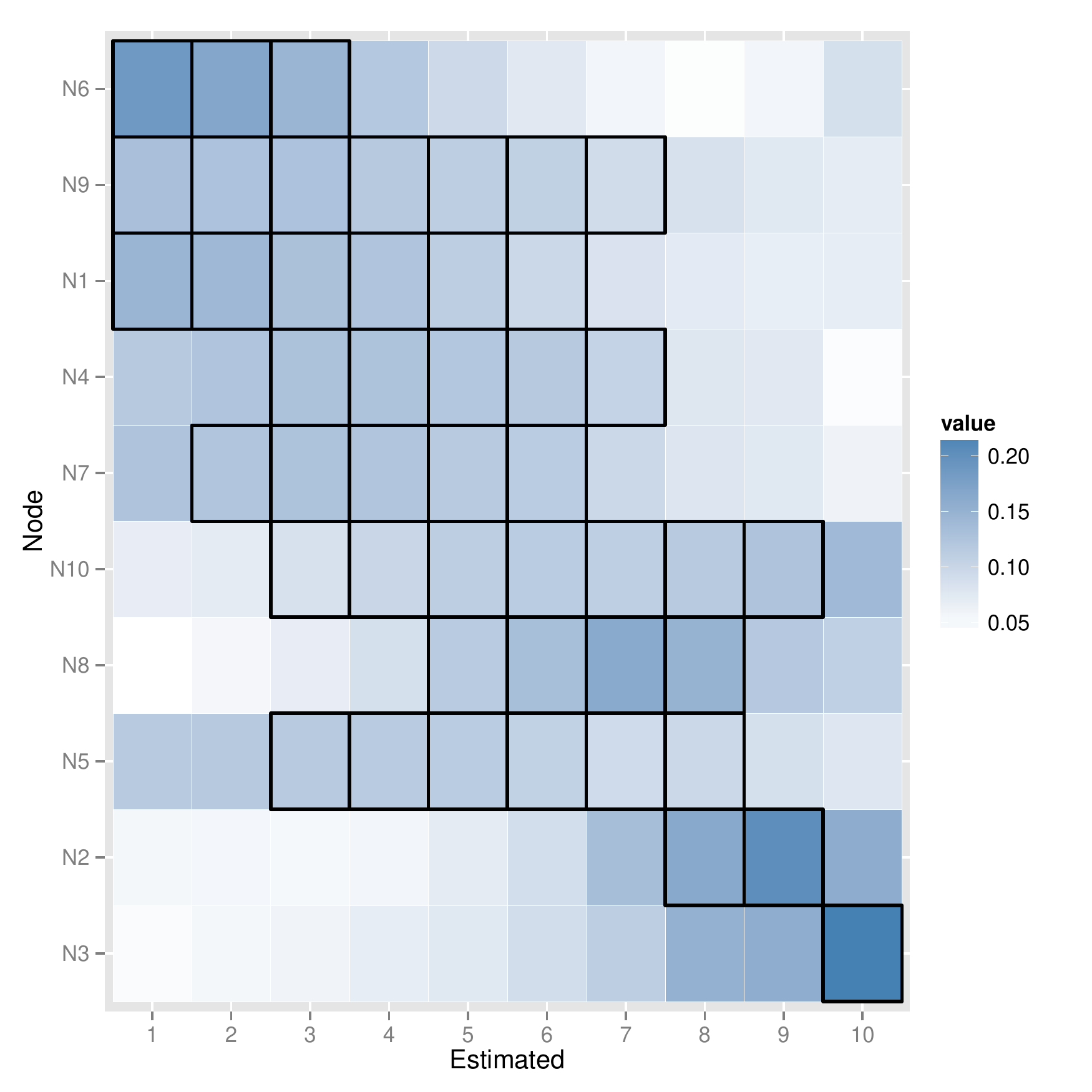}
\includegraphics[width=.40\textheight]{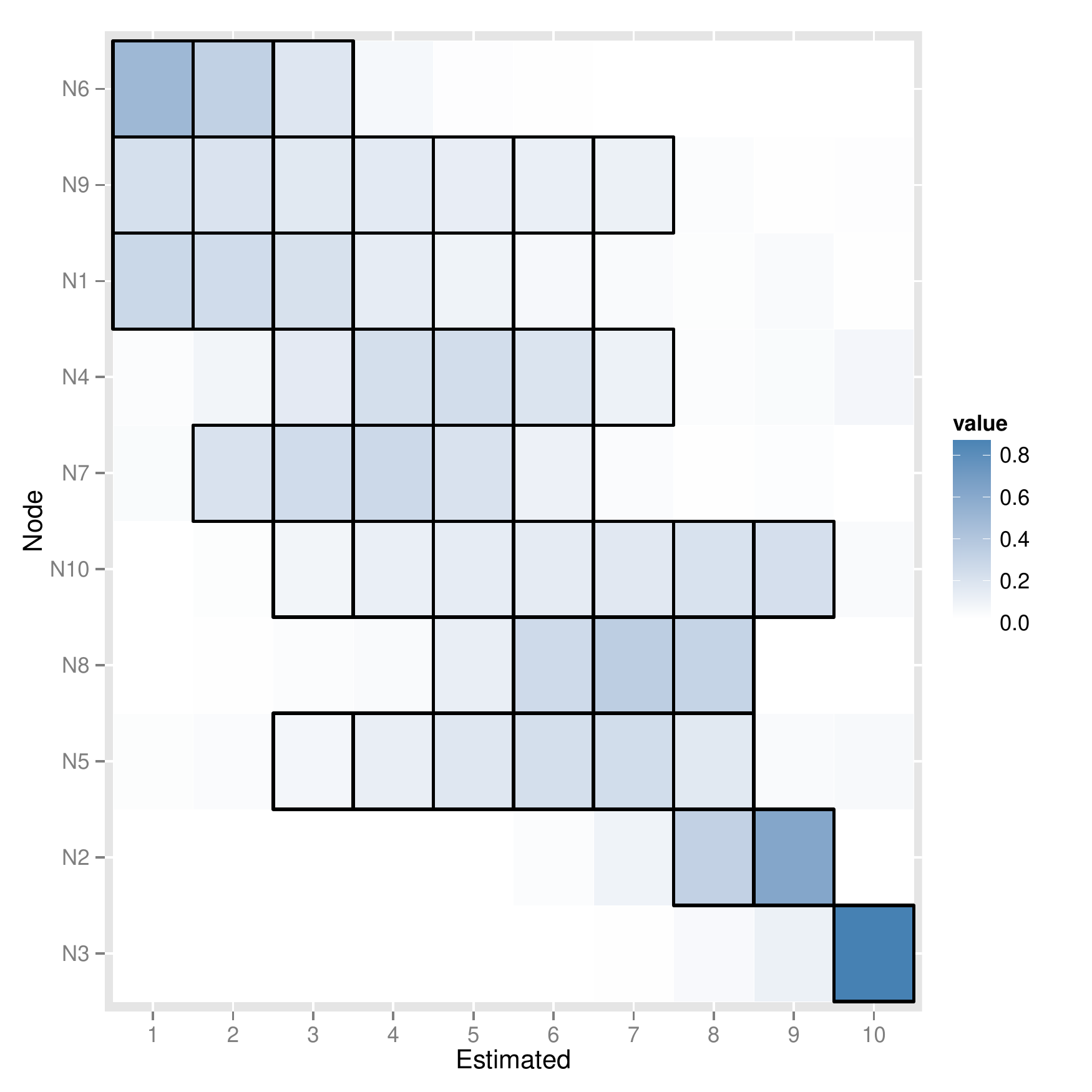}
\includegraphics[width=.40\textheight]{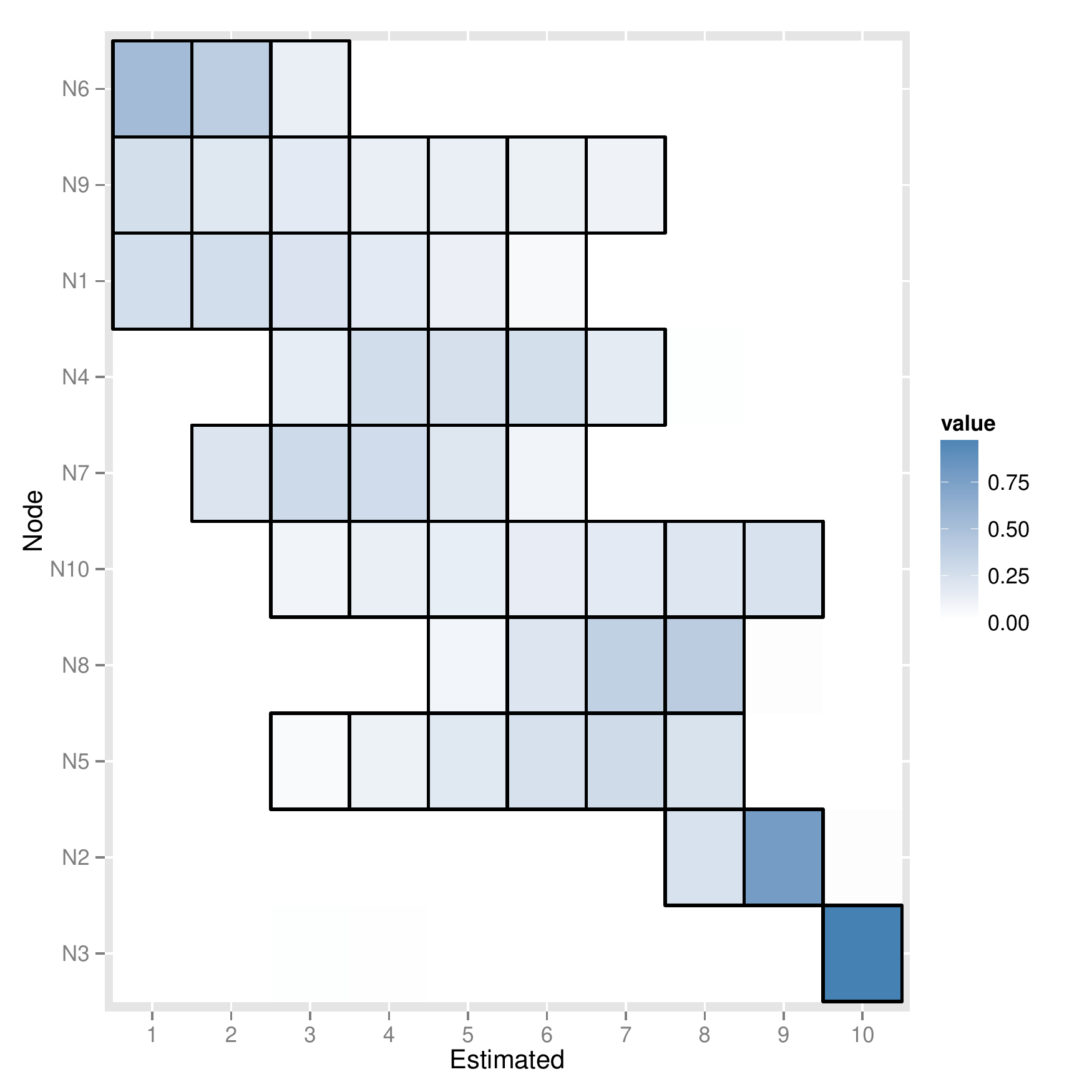}
\caption{{\bf $\sigma = 0.5$ (left), $\sigma = 0.1$ (middle), $\sigma = 0.01$ (right), mixed setting}. Posterior distribution of node orders, averaged over 100 simulations. The true node order (1 to 10) is represented in the rows, the estimated node orders in the columns, and the color of each square corresponds to the average proportion of iterations in which a given node was placed in a given position. As the node ordering is not unique for this DAG, true potential positions for each node are outlined in black.}
\end{figure}
\end{landscape}

\begin{landscape}
\begin{figure}[p]
\centering
\includegraphics[width=.40\textheight]{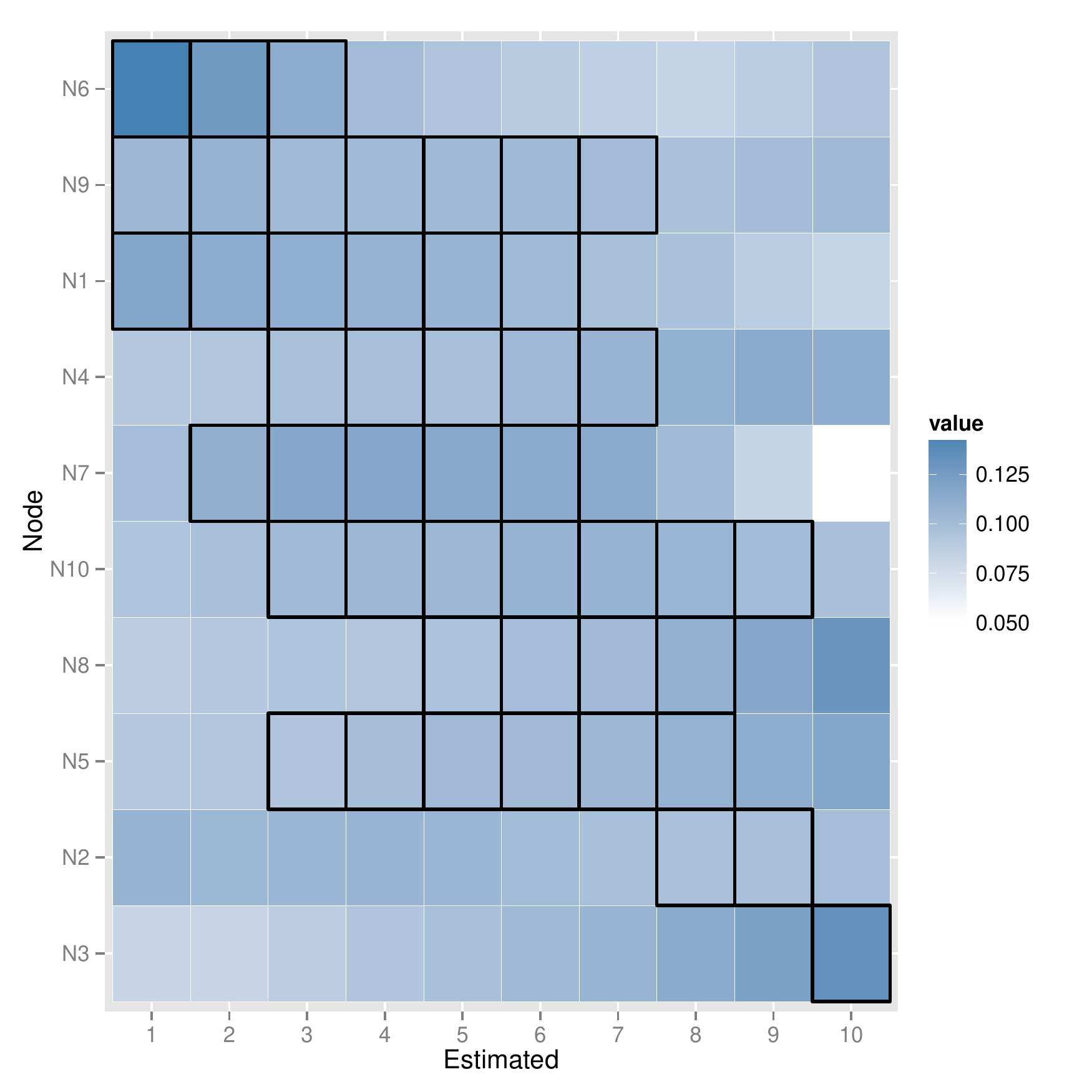}
\includegraphics[width=.40\textheight]{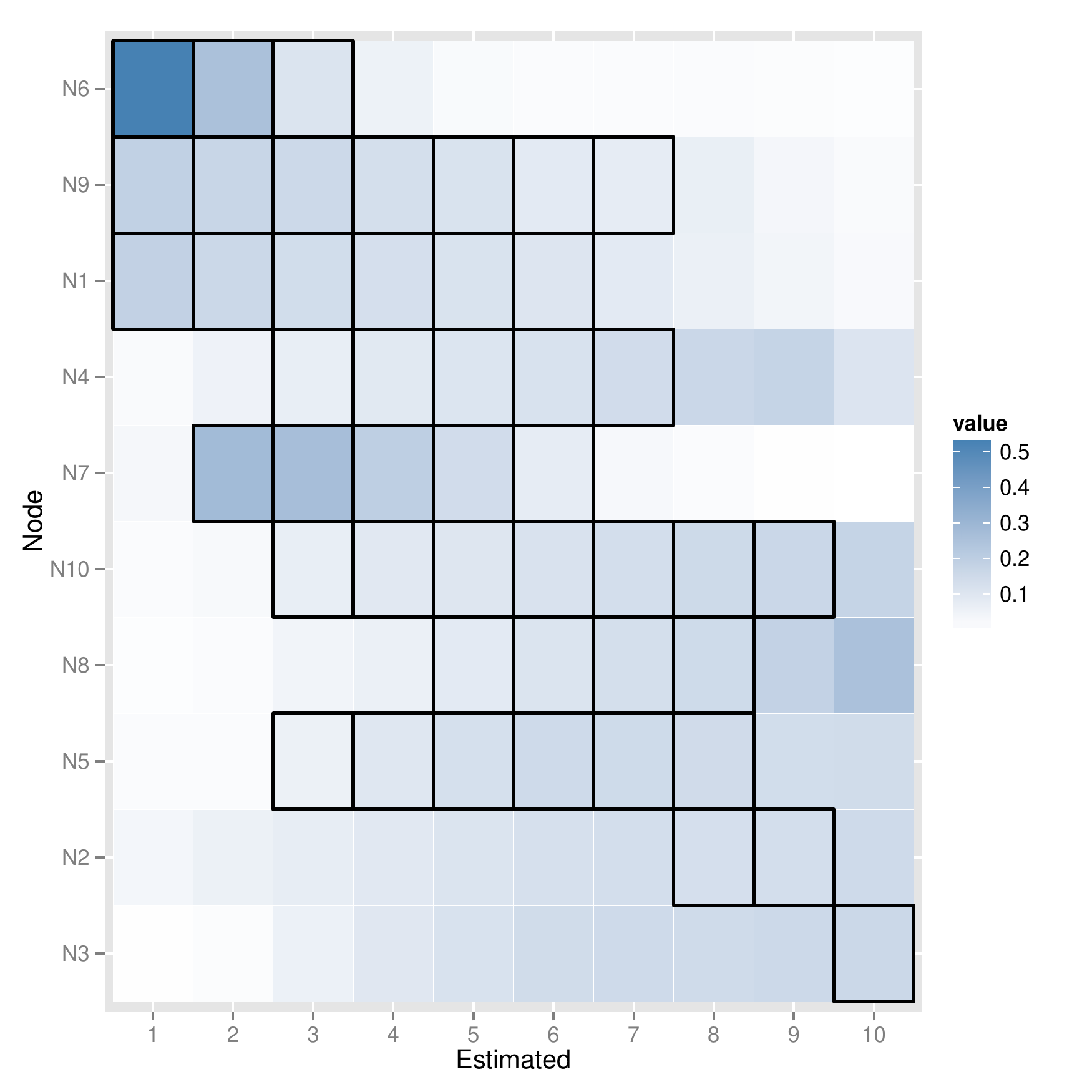}
\includegraphics[width=.40\textheight]{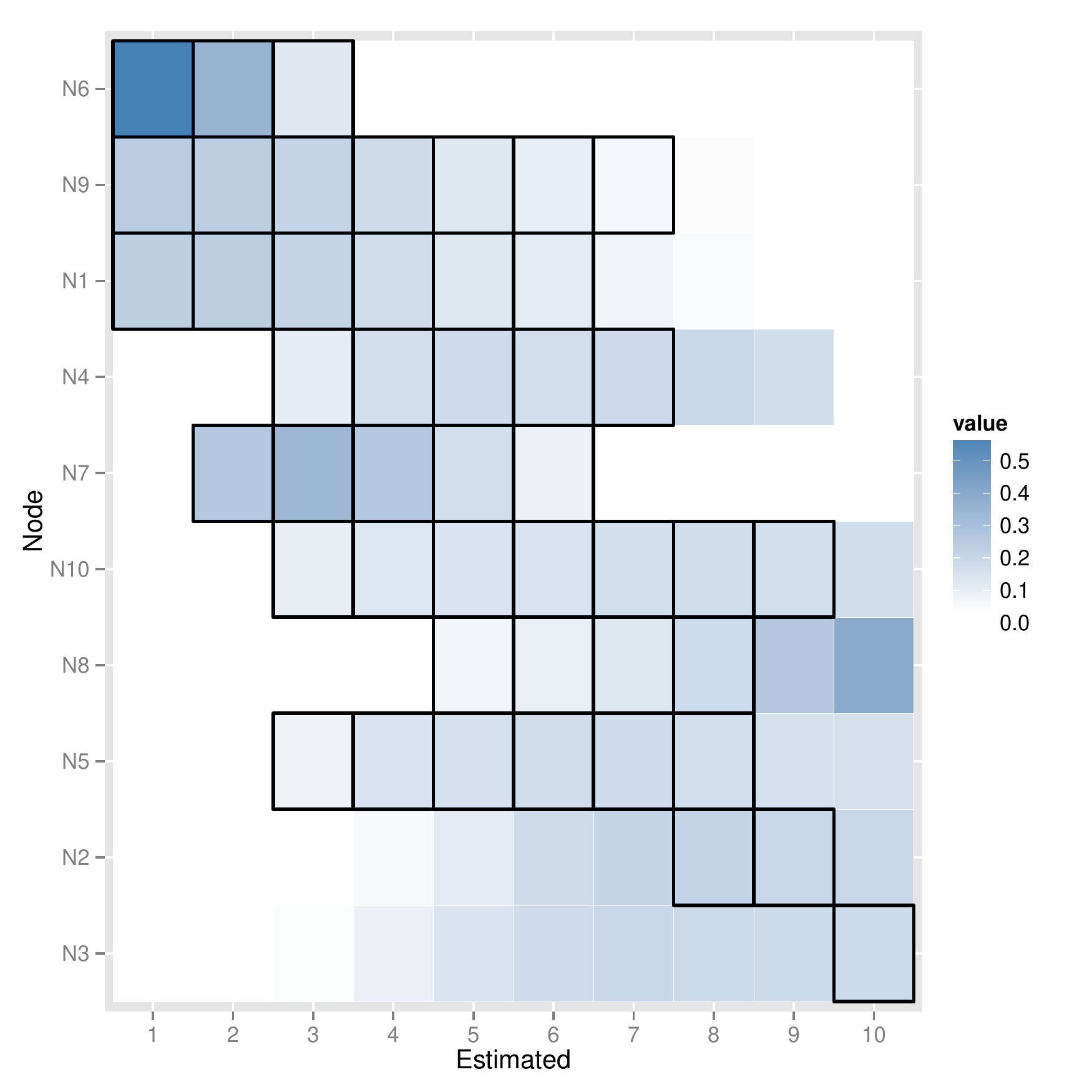}
\caption{{\bf $\sigma = 0.5$ (left), $\sigma = 0.1$ (middle), $\sigma = 0.01$ (right), partial KO setting}. Posterior distribution of node orders, averaged over 100 simulations. The true node order (1 to 10) is represented in the rows, the estimated node orders in the columns, and the color of each square corresponds to the average proportion of iterations in which a given node was placed in a given position. As the node ordering is not unique for this DAG, true potential positions for each node are outlined in black.}
\end{figure}
\end{landscape}

\begin{landscape}
\begin{figure}[p]
\centering
\includegraphics[width=.40\textheight]{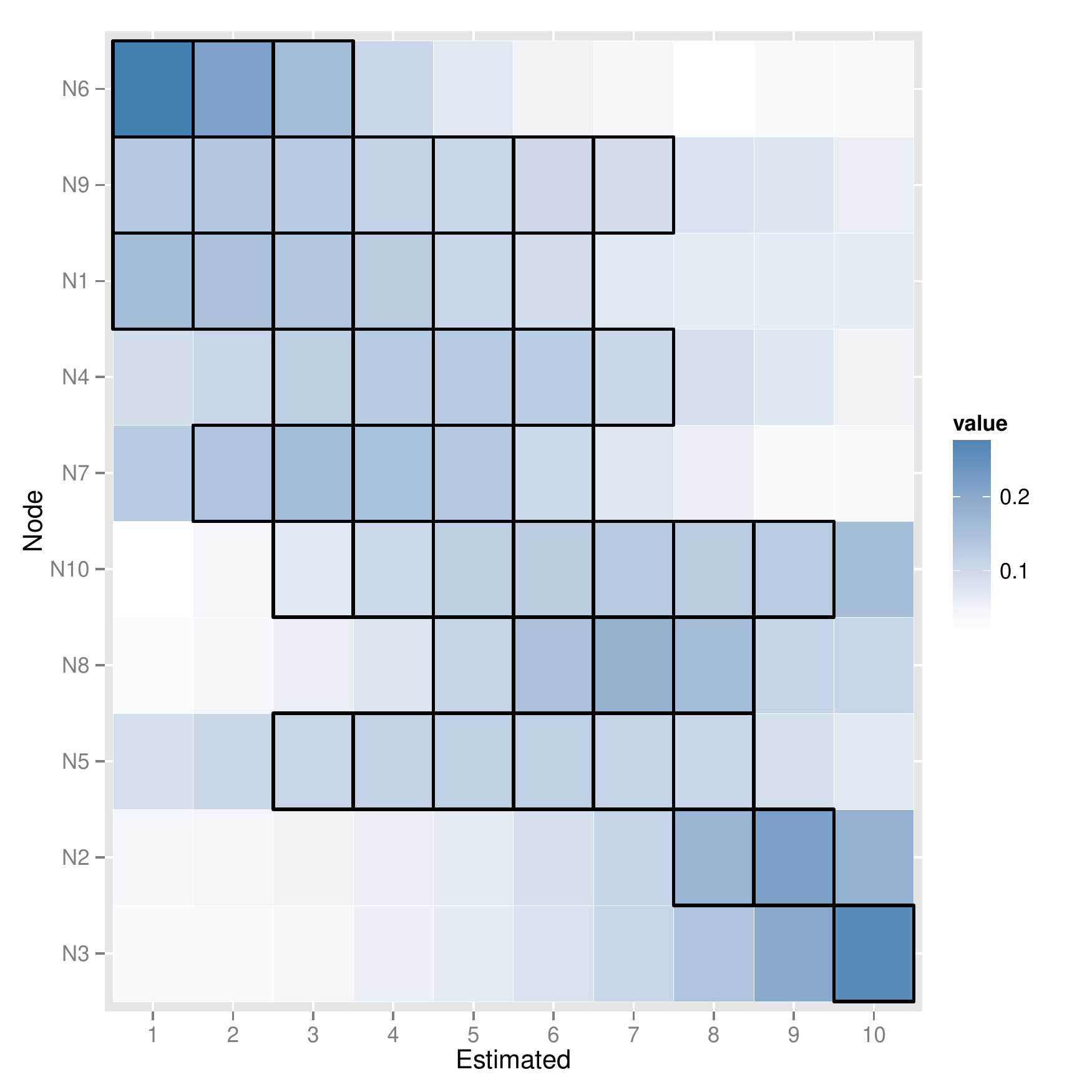}
\includegraphics[width=.40\textheight]{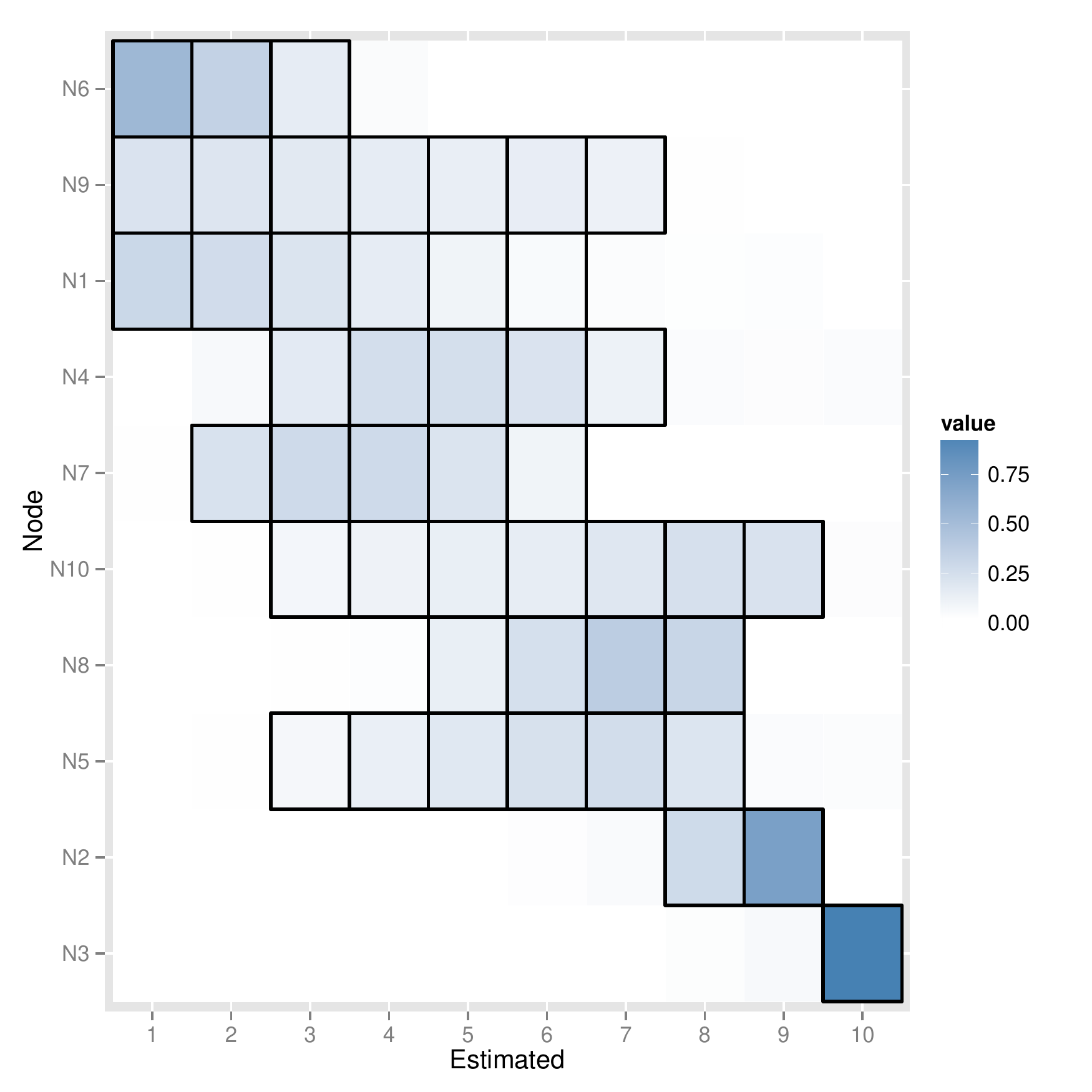}
\includegraphics[width=.40\textheight]{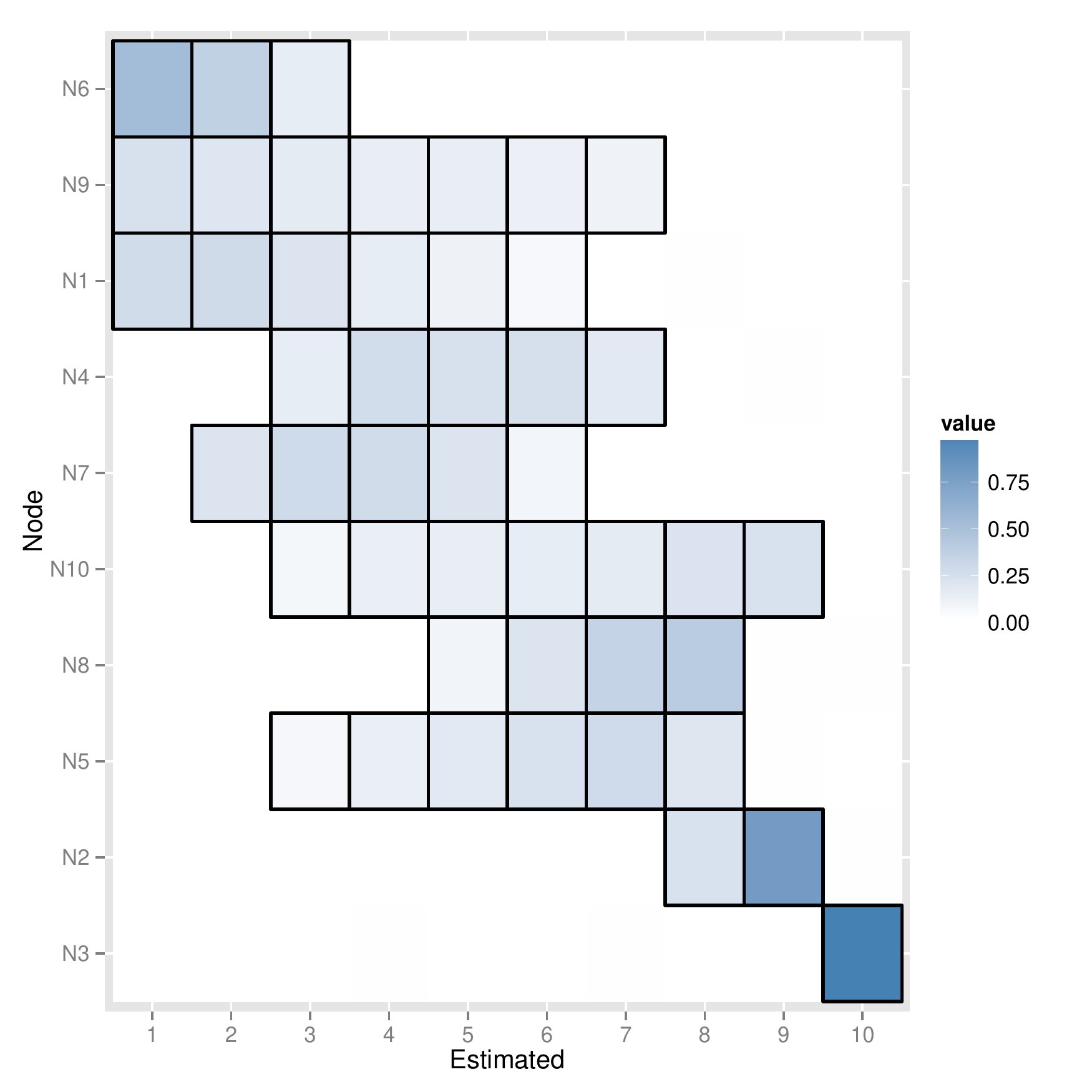}
\caption{{\bf $\sigma = 0.5$ (left), $\sigma = 0.1$ (middle), $\sigma = 0.01$ (right), multiple KO setting}. Posterior distribution of node orders, averaged over 100 simulations. The true node order (1 to 10) is represented in the rows, the estimated node orders in the columns, and the color of each square corresponds to the average proportion of iterations in which a given node was placed in a given position. As the node ordering is not unique for this DAG, true potential positions for each node are outlined in black.}
\end{figure}
\end{landscape}

\end{document}